\documentclass[9pt, conference]{IEEEtran}
\IEEEoverridecommandlockouts
\usepackage{cite}
\usepackage{amsmath,amssymb,amsfonts}
\usepackage{algorithmic}
\usepackage{graphicx}
\usepackage{textcomp}
\usepackage{xcolor}

\usepackage{bm, amssymb, booktabs, multirow,cite, url, xcolor, hyperref}
\usepackage{makecell, arydshln}
\usepackage{threeparttable}

\def\BibTeX{{\rm B\kern-.05em{\sc i\kern-.025em b}\kern-.08em
    T\kern-.1667em\lower.7ex\hbox{E}\kern-.125emX}}
\begin{document}
\bstctlcite{IEEEexample:BSTcontrol}

\title{Leveraging Self-Supervised Learning for  Speaker Diarization\\
\thanks{The work was supported by Czech Ministry of Interior projects Nos. VJ01010108 "ROZKAZ" and VK01020132 "112", and Horizon 2020 Marie Sklodowska-Curie grant ESPERANTO, No. 101007666. Computing on IT4I supercomputer was supported by the Czech Ministry of Education, Youth and Sports through the e-INFRA CZ (ID:90254).}
}

\author{\IEEEauthorblockN{Jiangyu Han,
Federico Landini, Johan Rohdin, Anna Silnova, Mireia Diez, and
Luk\'{a}\v{s} Burget}
\IEEEauthorblockA{Brno University of Technology, Speech@FIT, Czechia\\
\{ihan, landini, rohdin, isilnova, mireia, burget\}@fit.vut.cz
}}
\maketitle

\begin{abstract}
End-to-end neural diarization has evolved considerably over the past few years, but data scarcity is still a major obstacle for further improvements. 
Self-supervised learning methods such as WavLM have
shown promising performance on several downstream tasks, but their application on speaker diarization is somehow limited. 
In this work, we explore using WavLM to alleviate the problem of data scarcity for neural diarization training. We use the same pipeline as Pyannote and improve the local end-to-end neural diarization with WavLM and Conformer. Experiments on far-field AMI, AISHELL-4, and AliMeeting datasets show that our method substantially outperforms the Pyannote baseline and achieves new state-of-the-art results on AMI and AISHELL-4, respectively. In addition, by analyzing the system performance under different data quantity scenarios, we show that WavLM representations are much more robust against data scarcity than filterbank features, enabling less data hungry training strategies. Furthermore, we found that simulated data, usually used to train end-to-end diarization models, does not help when using WavLM in our experiments. 
Additionally, we also evaluate our model on the recent CHiME8 NOTSOFAR-1 task where it achieves better performance than the Pyannote baseline.
Our source code is publicly available at \url{https://github.com/BUTSpeechFIT/DiariZen}.
\end{abstract}

\vspace{0.4cm}

\section{Introduction}
\vspace{0.1cm}
Speaker diarizarion is the task of determining ``who spoke when'' in a multi-speaker recording. To better handle the overlapped speech, researchers have progressively moved from clustering-based approaches \cite{wang2018speaker, park2019auto, landini2022bayesian} to end-to-end neural diarization (EEND) \cite{fujita2019end, horiguchi20_interspeech, landini2024diaper, harkonen2024eend, chen2024attention}. Although the end-to-end methods demonstrate promising performance in certain scenarios, they struggle to process recordings with several speakers, i.e., more than four, especially when the input recording is long. Therefore, a few works have proposed integrating clustering-based diarization with EEND \cite{kinoshita2021integrating, kinoshita2021advances}. The general principle of this EEND-VC approach is to apply EEND on short chunks of the input recording first, then stitch together the local diarization results using speaker embeddings and clustering. However, most of the EEND-related methods do have a significant limitation:  they are extremely data hungry, typically requiring more than ten thousand hours of simulated data for model training \cite{landini2024diaper, harkonen2024eend, chen2024attention}. 
Moreover, due to the existing mismatch between simulated data and the target domain, further model adaptation is usually necessary. Such training costs make speaker diarization research arduous and complicated. 

\vspace{0.05cm}

To circumvent these limitations, one option is to apply the EEND model to short speech segments of a few seconds, assuming that only a small number of speakers are active. In such a constrained scenario, it is possible to train the model directly from scratch using the available real data. Along this line, following the EEND-VC principle, Pyannote \cite{bredin2023pyannote} achieves strong performance across different datasets. Compared to the EEND-VC methods in \cite{kinoshita2021integrating, kinoshita2021advances}, Pyannote uses much
shorter chunks (5s instead of 30s) for the local EEND processing.
Additionally, by introducing powerset loss \cite{plaquet2023powerset}, the Pyannote model can be further improved while eliminating the decision threshold of the EEND part, critical and sensitive for most neural diarization methods.

Another possible way to make diarization training less data hungry is to use self-supervised learning (SSL) methods, which attempt to make a single universal model applicable to a wide variety of tasks and domains \cite{mohamed2022self,baevski2020wav2vec, hsu2021hubert, chen2022wavlm}. Among them, WavLM \cite{chen2022wavlm} shows superior performance for speech processing. By including overlapped speech during model training, WavLM naturally has the potential to improve the model capabilities on the diarization task. In recent years, some diarization works managed to incorporate WavLM into their frameworks \cite{chen2022wavlm, delcroix2023multi, tawara2024ntt}. Although they got excellent diarization results, the existing methods do not make diarization training lightweight. The authors usually just replace log Mel-filterbanks with WavLM features, then train their models with thousands of hours of simulated data and fine-tune them with real target domain data.
In \cite{baroudi2023pyannote, kalda2024taltech}, WavLM has already been applied in the context of pyannote, but the analysis of their results was very limited.
In addition, to the best of our knowledge, no paper has yet analyzed the impact of either the quantity or the quality of training data on the final performance when using SSL methods. For example, when using a pre-trained WavLM, how much data is needed in order to train a diarization model to achieve reasonable performance? Considering WavLM has already been pre-trained on huge amounts of data, can we leverage it to make diarization training less data hungry? Do we really need simulated data when using WavLM? We believe these fundamental and unsolved questions are crucial to understanding how SSL models should be used for the speaker diarization task.

In this paper, we try to answer the questions above, aiming to make the training of speaker diarization more lightweight and less data hungry by means of SSL models. Due to the better performance of WavLM in comparison with other alternatives, we choose it for our experiments.
Following the Pyannote pipeline \cite{bredin2023pyannote, plaquet2023powerset}, we replace the original local EEND part with a new architecture composed of WavLM and Conformer \cite{gulati2020conformer}. To verify the effectiveness of our method, we use the far-field single-channel data from three public datasets, AMI \cite{carletta2005ami, kraaij2005ami}, AISHELL-4 \cite{fu2021aishell}, and AliMeeting \cite{yu2022m2met}, for system evaluation. Our experimental results demonstrate that the proposed method can significantly improve the Pyannote baseline and achieve strong performance on AMI and AISHELL-4 that is better than the current state-of-the-art. We compared the performance of our model with different amounts of real training data and showed how WavLM features are significantly superior to filterbanks in scenarios of data scarcity.
Even with just 14.4 hours of training data, our model reaches very competitive performance. Additionally, when using simulated data to train our model, the performance degrades, showing that simulated data is not necessarily beneficial when using SSL models. Finally, we also show improved performance on the NOTSOFAR-1 dataset \cite{vinnikov2024notsofar} in the context of the CHiME8 challenge.

\section{Methods}
\label{sec:method}
\subsection{Pyannote pipeline}

We follow the principle of EEND-VC \cite{kinoshita2021integrating, kinoshita2021advances} and use the Pyannote pipeline \cite{bredin2023pyannote, plaquet2023powerset}. During inference, given a long recording, it first splits the signal into overlapping short segments.
Then, EEND is applied on each segment to produce local diarization results.
After that, speaker embeddings are extracted on pure (without overlap) speech for each of the speakers in the segment. 
The extracted speaker embeddings are clustered using agglomerative hierarchical clustering (AHC) to find the across-segment speaker mapping. AHC is applied with the constraint that embeddings from the same segment do not end up in the same cluster. After mapping the speakers across segments, the local EEND decisions on frames of overlapping segments are aggregated by averaging the probabilities of the corresponding speakers on those frames.

\subsection{End-to-end neural diarization}

\begin{figure}[!htbp]
  \centering
  \includegraphics[width=8.8cm]{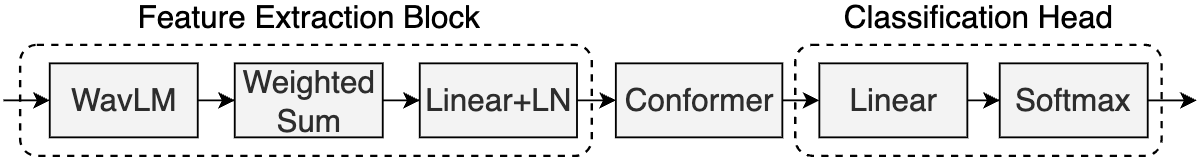}
  \caption{Proposed framework for local EEND module.}
  \label{fig:framework}
\end{figure}

The EEND part of Pyannote mainly consists of SincNet \cite{ravanelli2018speaker} and four bidirectional long short-term memory (LSTM) layers. 
Our proposed method closely follows the Pyannote pipeline with one major exception: we replace the EEND module with a model based on WavLM and Conformer shown in Figure \ref{fig:framework}.
Following the strategy proposed in SUPERB \cite{yang2021superb}, the WavLM outputs from each layer are weighted by learnable parameters and summed per frame to create a fused representation. Next, a linear layer and layer normalization (LN) are applied to transform the fused WavLM features before being passed to the Conformer. Then a linear layer and softmax are used to obtain the final classification outputs, which correspond to the powerset states in this paper.

We use the pre-trained WavLM Base+ \cite{chen2022wavlm} for all experiments. The outputs from its convolutional layers and 12 transformer encoders are combined using weighted sum.
The input and output dimensions for the following linear layer are 768 and 256, respectively. For Conformer, we follow the original architecture \cite{gulati2020conformer} while removing the positional encoding.
We use 4 Conformer blocks. For each block, the input and hidden dimensions for the feed-forward module are 256 and 1024, respectively; the number of attention heads for the multi-headed self-attention is 4; for the convolution module, the kernel size is set to 31. All dropout rates in the Conformer are set to 0.1. 
The powerset loss \cite{plaquet2023powerset} is used for model training, where we assume a maximum of 4 speakers and 2 overlapping speakers (plus one class for silence, which leads to 11 powerset classes). So the output dimension of the last linear layer is 11.
The numbers of parameters of WavLM and Conformer are 94.7 million and 6.1 million, respectively.

Additionally, given the widespread use of filterbank (Fbank) features in diarization tasks, we investigate their impact by substituting WavLM and the corresponding weighted sum with Fbank features. To utilize WavLM, we first consider freezing its parameters (WavLM-frozen), as in many studies \cite{chen2022wavlm, delcroix2023multi, tawara2024ntt}. Then following the joint optimization strategy \cite{peng2023attention}, we explore updating WavLM and other modules simultaneously (WavLM-updated).

\section{Experiments}
\label{sec:exp}
\vspace{-0.1cm}

\subsection{Datasets}
\vspace{-0.1cm}

We use the far-field single-channel data from the public datasets AMI \cite{carletta2005ami, kraaij2005ami}, AISHELL-4 \cite{fu2021aishell}, and AliMeeting \cite{yu2022m2met}, for system evaluation. Since AISHELL-4 lacks development data, we randomly select 10\% of the training data from each room for development, using the remaining data as the new training set. We then train our model using the combination of the three training sets, with their corresponding development sets combined and used for validation. Detailed per set and dataset information about the number of recordings, number of speakers, and duration, can be found in Table \ref{tab:dataset}. 


All our experiments in Section \ref{subsub: overal} and \ref{subsub: data_quantity} are conducted using only the compound real data. In Section \ref{subsub: data_simu}, we further explore the effects of simulated data. In Section \ref{subsub: nsf}, we also show the performance on the CHIME-8 NOTSOFAR-1 challenge  \cite{vinnikov2024notsofar}.

\begin{table}[tbp]
  \caption{Information of different datasets.}
  \label{tab:dataset}
 \setlength{\tabcolsep}{1mm}
  \centering
  \begin{tabular}{l | c c r | c c r | c c r}
    \hline
        \multirow{2}{*}{Dataset} & \multicolumn{3}{c|}{Train} & \multicolumn{3}{c|}{Dev} & \multicolumn{3}{c}{Test}\\
        & \#files & \#spk & \#hrs & \#files & \#spk & \#hrs & \#files & \#spk & \#hrs \\
    \hline
    AMI & 134 & 3-5 & 79.7 & 18 & 4 & 9.7 & 16 & 3-4 & 9.1 \\
    AISHELL-4 & 173 & 3-7 & 97.2 & 18 & 3-7 & 10.3 & 20 & 5-7 & 12.7 \\
    AliMeeting & 209 & 2-4 & 111.4 & 8 & 2-4 & 4.2 & 20 & 2-4 & 10.8 \\
    \hline
    Compound & 516 & 2-7 & 288.3 & 44 & 2-7 & 24.2 & 56 & 2-7 & 32.6 \\
    \hline
  \end{tabular}
  \vspace{-0.1cm}
\end{table}

\subsection{Configurations}
\vspace{-0.1cm}
\label{subsec: conf}
\subsubsection{Training}

We use our own code\footnote{https://github.com/BUTSpeechFIT/DiariZen} for model training. We split recordings into 8s segments with a hop size of 6s. The effective batch size is 64. 
For the filterbank experiments, we extract 80-dimensional Fbank features with 25 ms window size and 10 ms hop size. The optimizer is AdamW \cite{loshchilov2017decoupled}. 
When updating the parameters of WavLM and other modules simultaneously, we set the learning rate for WavLM to 1e-5 and 1e-3 for other parameters. In all other experiments, we set the learning rate to 1e-3. All models are trained up to 100 epochs, with early stopping applied if the validation loss does not decrease
for 10 consecutive epochs. We apply AutoClip \cite{seetharaman2020autoclip} to automatically choose a gradient clipping
threshold based on the 90-th percentile of gradient norms observed
during training. 

\subsubsection{Inference}

We use the Pyannote pipeline for inference. 
Model parameters are averaged 
over the last 5 checkpoints. Input recordings are split into 8s segments with a hop size of 0.8s.
We use ResNet34-LM\footnote{https://huggingface.co/Wespeaker/wespeaker-voxceleb-resnet34-LM} trained with the WeSpeaker toolkit \cite{wang2024advancing} on the VoxCeleb2 dataset \cite{chung2018voxceleb2} to extract the local speaker embeddings.
For AHC, the minimum and maximum number of clusters are set to 2 and 8, respectively. The minimum cluster size is 30. The clustering threshold applied to the cosine similarity is always set to 0.7. We use diarization error rate (DER) for system evaluation.
A macro-averaged DER is also reported to represent the overall performance across all datasets.

\begin{table*}[htbp]
  \caption{Far-field performance of DER across different datasets. Our best results are reported in bold.}
  \label{tab:overall}
  \centering
  \begin{threeparttable}
  \begin{tabular}{l| c | c c c | c c c}
    \hline
        \multirow{2}{*}{System} & \multirow{2}{*}{Features} & \multicolumn{3}{c|}{collar=0s} & 
        \multicolumn{3}{c}{collar=0.25s} \\
        & & AMI & AISHELL-4 & AliMeeting & AMI & AISHELL-4 & AliMeeting \\
    \hline
    Pyannote3 \cite{plaquet2023powerset} & SincNet & 22.0 & 16.9 & 23.3 & - & - & - \\
    \quad + fine-tuning \cite{plaquet2023powerset} & SincNet & 22.9 & 13.2 & 24.5 & 15.3 & 7.6 & 15.8 \\
    \hline
    Pyannote3 (baseline) & SincNet & 21.1 & 13.9 & 22.8 & 13.7 & 7.7 & 13.6 \\
    \hline
    \multirow{3}{*}{Proposed} & Fbank & 19.7 & 12.5 & 21.0 & 12.9 & 6.9 & 12.6 \\
    & WavLM-frozen & 17.0 & 11.7 & 19.9 & 10.9 & 6.1 & 12.0 \\
    & WavLM-updated & \textbf{15.4} & \textbf{11.7} & \textbf{17.6} & \textbf{9.8} & \textbf{5.9} & \textbf{10.2} \\
    \hline   
    State-of-the-art by August 2024 & - & 17.1 \cite{kalda2024pixit} & 12.2\tnote{*} & 13.4 \cite{harkonen2024eend} & 13.3 \cite{harkonen2024eend} & 7.6 \cite{plaquet2023powerset} & 6.1 \cite{harkonen2024eend} \\ 
    \hline
  \end{tabular}
  \begin{tablenotes}
    \footnotesize
    \item[*] 12.2 is reported in \url{https://github.com/pyannote/pyannote-audio/blob/0ea4c025ee048c36d74ccdb8b3f4939a27ad729b/README.md}
    \vspace{-0.2cm}
    \end{tablenotes}
  \end{threeparttable}
\end{table*}

\subsection{Results and discussion}
\subsubsection{Overall comparison}
\label{subsub: overal}

We show the overall performance of different systems in Table \ref{tab:overall}. First, the Pyannote3 results are copied from the original paper \cite{plaquet2023powerset}. The third line, Pyannote3 (baseline), represents the performance of the Pyannote model under our experimental setup as described in the section \ref{subsec: conf}. 
Note that here the training data of Pyannote3 (baseline) is a subset of the overall Pyannote3 \cite{plaquet2023powerset} training data.
We use the third line as the Pyannote baseline to ensure fair comparisons.

It is clear that our EEND part has better performance than the Pyannote baseline. Even with the traditional filterbank features, our model significantly outperforms the Pyannote model across the three datasets, suggesting that the Conformer is a better choice for the end-to-end module.
By utilizing WavLM, better performance can be obtained, especially for AMI and AliMeeting. These results suggest that WavLM can provide more powerful representations for the diarization task. Additionally, our model can be further improved by updating WavLM simultaneously with the Conformer training. 
Compared to the Pyannote3 baseline, when considering no forgiveness collar, our best model achieves relative DER reductions on AMI, AISHELL-4, and AliMeeting of 26.6\%, 15.8\%, and 22.8\%, respectively.

We also collected the best results available in the literature. As shown in Table \ref{tab:overall}, our results on AMI and AISHELL-4 surpass the best published ones at the time of writing. For AliMeeting, our performance is still far away from the best numbers. One possible reason is that AliMeeting only contains 2 to 4 speakers per session, which may be easier to model in a fully end-to-end manner \cite{harkonen2024eend}. 
More analysis on this needs to be done in the future.

\subsubsection{Effects of data quantity}
\label{subsub: data_quantity}
Since WavLM has already been pre-trained using large volumes of data, it might make the diarization training less data hungry. Although this hypothesis is intuitive, there has not been experimental support in the literature so far. To investigate the effect of data quantity, we randomly select subsets from AMI, AISHELL-4, and AliMeeting using  75\%, 50\%, 25\%, and 5\% of the total, respectively. These subsets are then combined to create new training datasets for each corresponding data size. We ensure that the smaller sets are always subsets of the larger ones.

\begin{figure}[tbp]
  \centering
  \includegraphics[width=8.8cm]{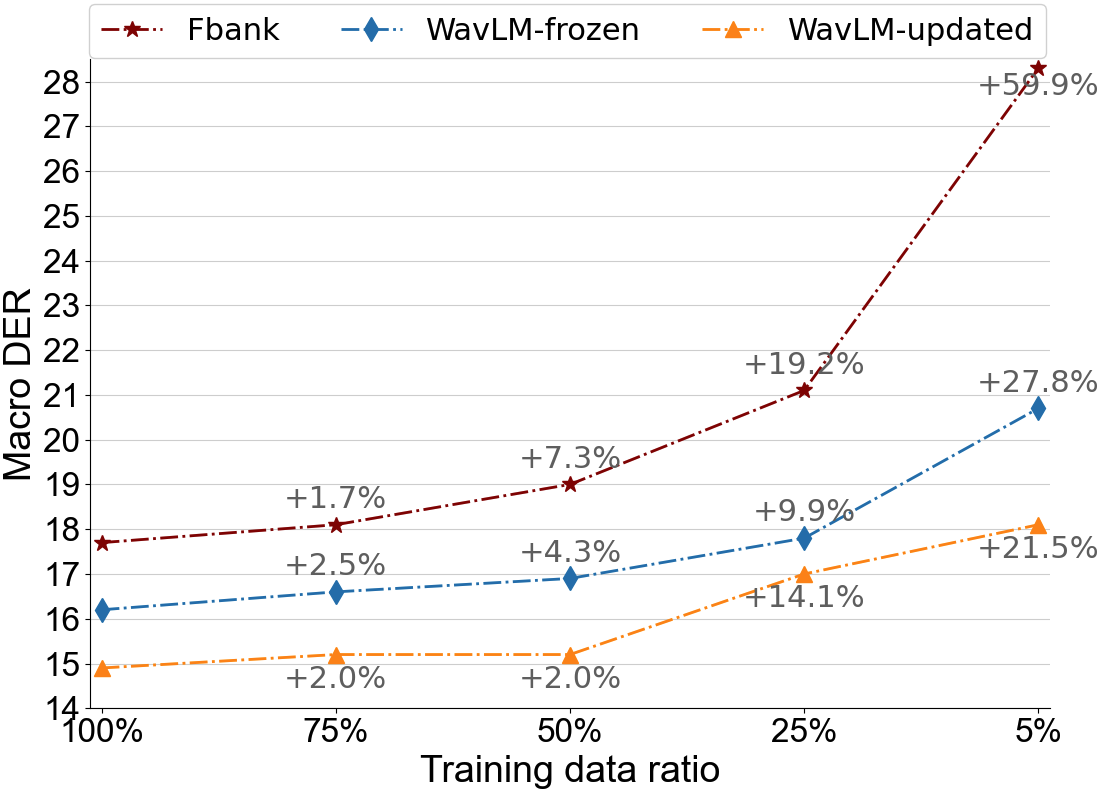}
  \caption{Macro-averaged DER when using different proportions of the compound set for training. The gray text indicates the relative performance degradation compared to using the entire compound set.}
  \label{fig:data_ratio}
  \vspace{-0.5cm}
\end{figure}

Detailed results can be found in Figure \ref{fig:data_ratio}, where we show the macro-averaged performance across the three datasets when using Fbank, WavLM-frozen, and WavLM-updated features for different training data sizes. As expected, the WavLM features have much better performance than Fbank under each condition. When reducing the available training data, all approaches tend to degrade. However, the effect is more attenuated for WavLM-based systems leading to better results overall. Such a phenomenon indicates that the SSL-based approach is more robust to data scarcity.
For an extreme case where only 5\% of data is available, the relative DER degradation is about 60\% when using Fbank features, while for the WavLM-based systems, it is less than 30\%. This puts into perspective how much effect the pre-trained SSL model can have on diarization tasks.

In Table \ref{tab:data_ratio}, we show the DER results of WavLM-updated for each dataset separately. Surprisingly, even with 5\% (14.4 hours) training data, our WavLM-updated system still performs better than the Pyannote baseline which is trained using the whole compound set. Therefore, based on our experiments, we conclude that our method, leveraging the pre-trained WavLM can significantly reduce the amount of data required for neural diarization training.

\subsubsection{Effects of data simulation}
\label{subsub: data_simu}
Data simulation is an essential part for EEND methods, yet creating appropriate simulations has always been challenging \cite{yamashita2022improving, landini2022simulated, landini2023multi}. By utilizing the strengths of SSL models like WavLM, commonly pre-trained on extensive data, the necessity of further data simulation may be diminished. To explore the effects of data simulation when using WavLM, we generate 1000 hours of wide-band simulated conversations\footnote{https://github.com/BUTSpeechFIT/EEND\_dataprep} from LibriSpeech \cite{panayotov2015librispeech}, with each recording containing 2 to 4 speakers at a ratio of 1:1:2.

\begin{table}[tbp]
  \caption{Performance of DER when training the model with different data ratios. The WavLM-updated setup is always applied.}
  \label{tab:data_ratio}
  \centering
  \begin{tabular}{l| c | c c c | c}
    \hline
       \multirow{2}{*}{Ratio} & \multirow{2}{*}{Hours} & \multicolumn{3}{c|}{collar=0s} & \multirow{2}{*}{Macro} \\
       & & AMI & AISHELL-4 & AliMeeting & \\ 
    \hline
    100\% & 288.3 & 15.4 & 11.7 & 17.6 & 14.9 \\
    75\% & 216.2 & 15.9 & 12.1 & 17.7 & 15.2 \\
    50\% & 144.1 & 16.1 & 12.5 & 17.0 & 15.2 \\
    25\% & 72.1 & 18.1 & 12.5 & 20.4 & 17.0 \\
    5\% & 14.4 & 19.7 & 12.8	& 21.7	& 18.1 \\
    \hline   
  \end{tabular}
  \vspace{-0.2cm}
\end{table}

\begin{figure}[tbp]
  \centering
  \includegraphics[width=8.8cm]{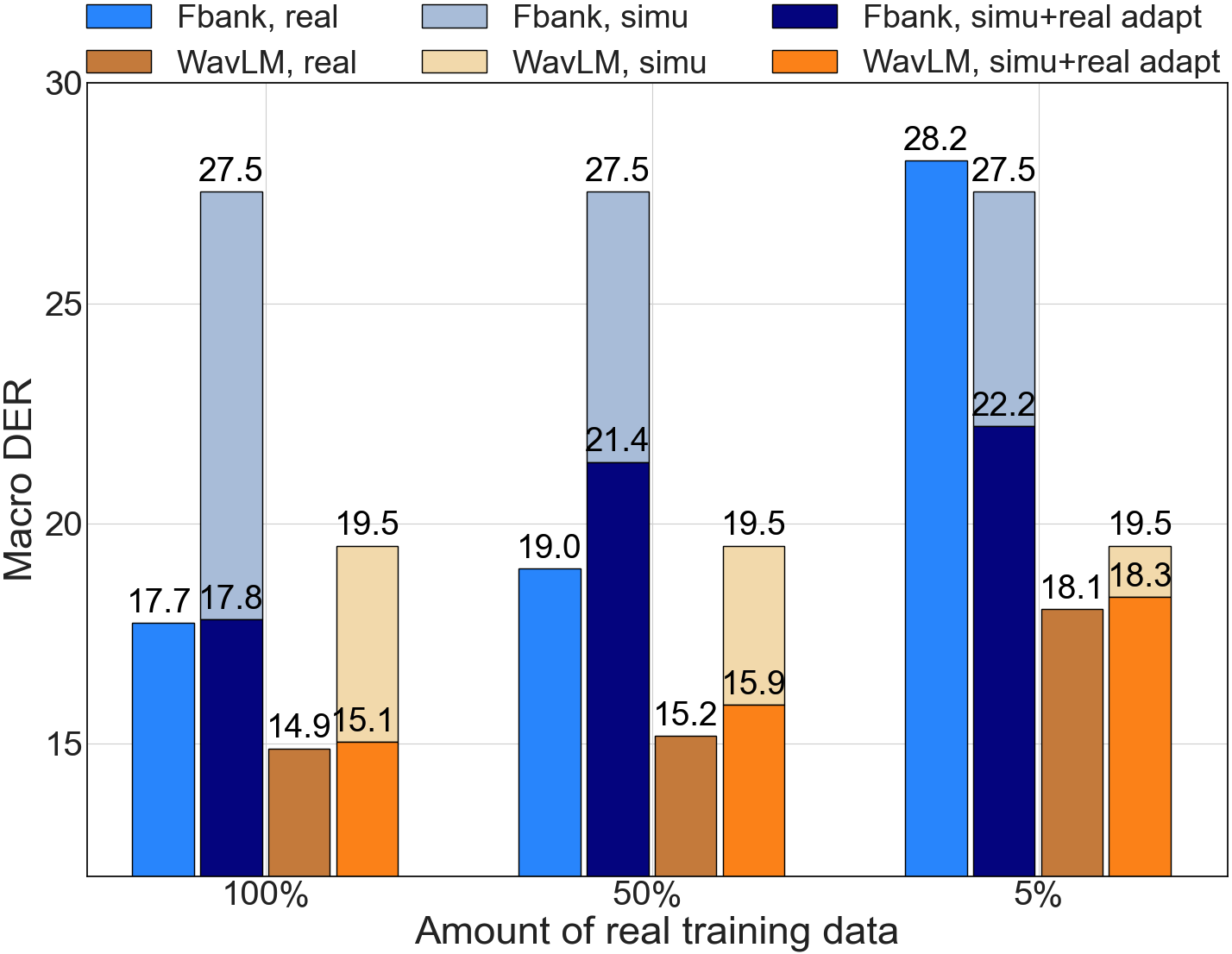}
  \caption{Macro-averaged performance of different training setups. The overlapping bars correspond to the results before and after adaptation to the corresponding training set to the current condition.}
  \label{fig:simu_real_bar}
  \vspace{-0.45cm}
\end{figure}

Our results can be found in Figure \ref{fig:simu_real_bar}, where we present the macro-averaged performance on AMI, AISHELL-4, and AliMeeting datasets for Fbank and WavLM features.
For each setup, we show the performance of 
training the model from scratch with subsets of the compound real data (Fbank, real; WavLM, real), 
training the model using simulated data alone (Fbank, simu; WavLM, simu), and training the model with simulated data then fine-tuning the model with the corresponding subsets of the compound real data (Fbank, simu+real adapt; WavLM, simu+real adapt). Note that the WavLM parameters are always initialized from the pre-trained model. 
When training the WavLM-based system with simulated data, the WavLM parameters are kept frozen since we have observed that updating them degrades the performance on the development set.
For the remaining scenarios, we update WavLM and the other modules simultaneously. 

As we can see, the real data always allows for better results than the simulated data. Except for the Fbank system in 5\% data condition, the simulated data generally has a negative impact. Our experiments show that using simulated data for WavLM-updated system does not help. When enough real data is available, it is difficult to obtain further benefits from the simulated data due to the domain mismatch. This is in contrast with the telephony scenario, usually studied in the context of EEND works \cite{fujita2019end, horiguchi20_interspeech, landini2024diaper, harkonen2024eend, chen2024attention}. Generating simulated data of good quality in such a case is possible due to the relatively simple acoustic conditions and abundant availability of telephone recordings. However, in other scenarios, especially for the far-field data as studied in this work, generating simulated training data that matches the evaluation conditions is more complex. While generating suitable simulated data might be possible, it would be costly and require training the model on thousands of hours. Alternatively, utilizing SSL models, like WavLM in our case, seems to have the potential to allow training directly on real data and achieve strong performance.

\subsubsection{Performance on NOTSOFAR-1}
\label{subsub: nsf}

Initially, our method was designed for the CHIME-8 NOTSOFAR-1 challenge  \cite{vinnikov2024notsofar}. However, the NOTSOFAR setting was somewhat limited for further comparisons. To verify the effectiveness of our method, we conducted experiments and analyses based on the well-known and established datasets as described earlier. Here we also provide our results on the NOTSOFAR-1 task.
This dataset comprises 315 unique meetings, each lasting on average 6 minutes with 4-8 attendees, featuring authentic, multi-participant English conversations recorded in about 30 different conference rooms at Microsoft offices. Each
meeting is recorded by up to 7 commercial far-field array devices. 


In this paper, we only consider the single-channel track, where each recording either comes from one channel far-field device or is a post-processed signal from an array device. Our results can be found in Table \ref{tab:nsf}, where we train models using \verb"train_sc" dataset and take \verb"dev_sc" dataset for validation and evaluation. Here, all Pyannote3 experiments are conducted using the original Pyannote code\footnote{https://github.com/pyannote/pyannote-audio}. In the first line, we show the performance of the pre-trained Pyannote model\footnote{https://huggingface.co/pyannote/segmentation-3.0}. Then we fine-tune (FT) that model or train it from scratch (TFS) with the NOTSOFAR \verb"train_sc" data. 
In comparison,  our WavLM-updated system clearly achieves more competitive performance, even when the non-WavLM modules are trained from scratch with the limited available real data.

\begin{table}[tbp]
  \caption{Performance on the single-channel NOTSOFAR development data. FT means fine-tuning. TFS means training from scratch.}
  \label{tab:nsf}
  \centering
  \begin{tabular}{l| c | c | c c c c }
    \hline
       \multirow{2}{*}{System} & \multirow{2}{*}{FT} & \multirow{2}{*}{TFS} & \multicolumn{4}{c}{collar=0.25s} \\
       & & & DER & Miss & FA & Confusion \\ 
    \hline   
    \multirow{3}{*}{Pyannote3 \cite{plaquet2023powerset}} & - & - & 17.3 & 7.6 & 2.6 & 7.1 \\
        & \checkmark & - & 13.8 & 4.0 & 2.9 & 6.9 \\
        & - & \checkmark & 18.0 & 5.0 & 3.8 & 9.2 \\
    \hline
    WavLM-updated & - & \checkmark & 12.4 & 3.2 & 2.6 & 6.6 \\
    \hline
  \end{tabular}
  \vspace{-0.2cm}
\end{table}

\subsubsection{Inference budget}
To analyze our approach from a practical perspective, we show the inference memory and real time factor (RTF) of the EEND part for different systems in Table \ref{tab:infer_budget}. The test recording is half an hour long and comes from the AMI corpus.
As we can see, Pyannote3 has the fastest inference speed. When using a GPU, our methods are slightly slower than the Pyannote model, but still orders of magnitude faster than real-time. However, when using a CPU, the WavLM-based system is considerably slower. 
For the rest of the Pyannote pipeline, the inference budget for clustering is negligible, while the speaker embedding extraction is very costly. 
When extracting speaker embeddings on a GPU, all systems have RTFs of 0.02. However, when using a CPU, the corresponding RTF for embedding extraction of Pyannote3, our Fbank, and WavLM-based systems are 2.79, 2.92, and 2.42, respectively.
Here the difference in RTFs mainly comes from the different frame rates of different systems. 
While the RTF for Pyannote3 is much faster on CPU than for the proposed method, the overall RTF is still dominated by the embedding extraction step common to all methods. Therefore, the global RTFs for the different methods are not so different.

\begin{table}[htbp]
  \caption{Inference budget when using GPU or CPU. The GPU used is NVIDIA RTX A5000. The CPU used is Intel(R) Xeon(R) CPU E5-2640 v4 @ 2.40GHz. The inference batch size is 32.}
  \label{tab:infer_budget}
  \setlength{\tabcolsep}{1.6mm}
  \centering
  \begin{tabular}{l| c | c c | c c}
    \hline
        \multirow{2}{*}{System} & \multirow{2}{*}{Features} & \multicolumn{2}{c|}{GPU usage} & \multicolumn{2}{c}{CPU usage} \\
        & & Memory & RTF & Memory & RTF \\ 
    \hline
    Pyannote3 & SincNet & 1.0 GB  & 0.01 & 1.0 GB & 0.05 \\
    \hline
    \multirow{3}{*}{Proposed} & Fbank & 3.1 GB & 0.01 & 1.0 GB & 0.35 \\
    & WavLM-frozen & 5.8 GB & 0.02 & 3.8 GB & 2.36 \\
    & WavLM-updated & 5.8 GB & 0.02 & 3.8 GB & 2.36 \\
    \hline   
  \end{tabular}
  \vspace{-0.1cm}
\end{table}

\section{Conclusion}
\label{sec:conclusion}
\vspace{-0.1cm}

In this study, we proposed to use self-supervised learning models to alleviate the problem of data scarcity for neural speaker diarization. We used WavLM and followed the Pyannote pipeline to conduct experiments and analysis. The results on far-field AMI, AISHELL-4, and AliMeeting datasets prove that our method achieves strong performance. 
Moreover, we conclude that using the pre-trained WavLM can greatly reduce the data quantity required for neural diarization training.
In addition, by exploring the effects of data simulation, we found that simulated data is not needed in this framework. Besides, we also provide comparisons on the NOTSOFAR-1 dataset. Finally, we show the inference budget of different systems from a practical perspective. Our code is open-sourced to ensure reproducibility. 

\bibliographystyle{IEEEtran}
\bibliography{refs}
\end{document}